\documentclass[ACS]{WileyNJD-v1}

\articletype{Article Type}%

\usepackage[utf8x]{inputenc}
 \usepackage[T1]{fontenc}
\usepackage{diagbox}
\usepackage{amsfonts}
\usepackage{amsmath}
\usepackage{amsthm}
\usepackage{indentfirst}
\usepackage{microtype}
\usepackage[squaren]{SIunits}
\usepackage{booktabs}
\usepackage{graphicx}
\usepackage{multirow}
\usepackage{wallpaper}

\usepackage{braket}
\usepackage{caption}
 \makeatletter
 \g@addto@macro\normalsize{%
   \setlength\abovedisplayskip{15pt}
   \setlength\belowdisplayskip{15pt}
   \setlength\abovedisplayshortskip{15pt}
   \setlength\belowdisplayshortskip{15pt}
 }

\makeatother

\begin{document}
\title{Local structure in dense hydrogen at the liquid-liquid phase transition by Coupled Electron-Ion Monte Carlo}

\author[1,2]{Carlo Pierleoni*}

\author[3,4]{Markus Holzmann}

\author[5]{David M. Ceperley}

\authormark{Carlo Pierleoni \textsc{et al}}

\address[1]{\orgdiv{Department of Physical and Chemical Sciences}, \orgname{University of L'Aquila}, \orgaddress{\country{Italy}}}

\address[2]{\orgdiv{Maison de la Simulation, CEA, CNRS, Univ. Paris-Sud, UVSQ}, \orgname{Universit\'e Paris-Saclay}, \orgaddress{\country{91191 Gif-sur-Yvette, France}}}

\address[3]{\orgdiv{LPMMC, UMR 5493 of CNRS}, \orgname{Universit\'e Grenoble Alpes}, \orgaddress{\country{France}}}

\address[4]{\orgdiv{Institut Laue-Langevin}, \orgname{BP 156, F-38042 Grenoble Cedex 9}, \orgaddress{\country{France}}}

\address[5]{\orgdiv{Department of Physics}, \orgname{University of Illinois Urbana-Champaign}, \orgaddress{\state{Illinois}, \country{USA}}}

\corres{*Carlo Pierleoni, \email{carlo.pierleoni@aquila.infn.it,carlo.pierleoni@cea.fr}}



\abstract[Summary]{We present a study of the local structure of high pressure hydrogen around the liquid-liquid transition line based on results from the Coupled Electron-Ion Monte Carlo method. We report results for the Equation of State, for the radial distribution function between protons $g(r)$ and results from a cluster analysis to detect the possible formation of stable molecular ions beyond the transition line, as well as above the critical temperature. We discuss various estimates for the molecular fraction in both phases and show that, although the presence of $H_3^+$ ions is suggested by the form of the $g(r)$ they are not stable against thermal fluctuations.}


\maketitle

\section{Introduction}
Hydrogen under pressure has an interesting phase diagram with several experimentally detected crystalline structures at low temperature \cite{McMahon2012a,Rillo2017}. In the fluid phase at higher temperatures, experiments indicate the presence of a discontinuous increase in reflectivity which is interpreted as the insulator-to-metal transition \cite{Zaghoo2016,Ohta2015,Knudson2015} although a marked disagreement between results from different methods (static and dynamic compression) is observed. The theoretical scenario is that, compressing molecular hydrogen at constant temperature and below some still to be determined critical temperature, a first order phase transition is observed with a small density discontinuity similar to the textbook gas-liquid transition \cite{Morales2010,Lorenzen2010,Morales2013liquid,Pierleoni2016}. The weakly first order character of the transition and the absence of detectable metastable states makes possible the direct detection of the transition pressure avoiding the computation of the free energies of the coexisting phases. Ab-initio theories at different levels of accuracy provide the same qualitative scenario but the location of the transition line depends on the details of the theory, the most accurate calculation to date being the one provided by the Coupled Electron-Ion Monte Carlo method \cite{Pierleoni2016}. At the transition the system becomes suddenly conductive, a result obtained so far within the single electron picture of DFT, providing evidence that the density discontinuity is related to metallization. 
Less clear is the link with the molecular dissociation, a concept that originated from the chemical picture and which is rather qualitative. 
In particular, it is not clear if metallization occurs in a partially dissociated molecular environment or if dissociation is also a discontinuous process which occurs simultaneously with metallization. 
This question has been investigated in the past by several authors \cite{Vorberger2007a,Holst2008} both using the information from the proton-proton pair correlation function and by employing a cluster analysis and the concept of residence time. The first method is rather qualitative while the latter uses informations about the nuclear dynamics which are thought to provide a better characterization.

Note that all ab-initio theories are based on the Born-Oppenheimer (BO) approximation ignoring electron dynamics and therefore also the true dynamics of the dissociation/recombination process. Within the BO approximation, the electrons only provide a potential energy surface for the nuclear motion/sampling of the nuclear configurations. In these conditions, the concept of bounded or unbounded pairs of ions is probabilistic and can be addressed without analyzing the dynamics but only using the sampling of the nuclear configuration space as provided by the CEIMC.

In this paper we revisit our recent CEIMC study\cite{Pierleoni2016} providing a more complete account of the results both for the thermodynamics and for the structure of the liquid around the transition line. In particular, we try to quantitatively answer the question of whether we are able to detect the signature of stable aggregates (either neutral molecules, or molecular ions $H_2^+$ and $H_3^+$) even beyond the transition pressure along the various isotherms. In agreement with previous findings \cite{Vorberger2007a} we find that the information provided by the proton-proton pair correlation is misleading, suggesting a much larger molecular character, while the cluster analysis gives a more precise indication. On this basis we confirm that the observed thermodynamic discontinuity signals both a sudden metallization and molecular dissociation although a more quantitative analysis of molecular stability remains to be performed. This is at variant with a more gradual molecular dissociation observed in BOMD simulations with various functionals\cite{Morales2010,Knudson2015}, a result however inferred by the analysis of the pair correlation function only.

The paper is organized as follows. In the next section we review very briefly the CEIMC method. The following section is devoted to a more extensive presentation of CEIMC results than in ref. \cite{Pierleoni2016}, in particular of the pair correlation functions across the LLPT. The following section discuss the results of our cluster analysis and our conclusions.

\section{Method}
The Coupled Electron-Ion Monte Carlo method is a Quantum Monte Carlo ab-initio method for a system of electrons and nuclei\cite{Pierleoni2006,McMahon2012a}. Electrons are assumed to be in their ground state, while nuclei are at finite but low enough temperatures for the adiabatic approximation to be applicable. 
For hydrogen, the only system considered so far, the non-relativistic Hamiltonian comprises nuclear and electron kinetic energy operators and the Coulomb potential between any pair of particles. At given nuclear positions, the electronic energy is obtained by ground state Quantum Monte Carlo, either variational or with a projection operator within the fixed node approximation. Nuclear configurations are then sampled from the equilibrium Boltzmann distribution, either for classical point particles or for quantum nuclei within the Path Integral formalism. Considering non-classical nuclear statistics, although possible, has not yet been attempted. 

The most important ingredient of CEIMC is the many electron wave function which sets the accuracy of the solution. We use the Slater-Jastrow form with a single determinant per spin type. Those determinants are formed with single electron orbitals obtained from a self-consistent Density Functional Theory (DFT) solution with the PBE xc approximation. A backflow transformation of electron coordinates in quasi-particle coordinates is further applied. The Jastrow part of the wave function comprises single body (proton-electron), two-body (electron-electron) and three-body terms.
The analytical form of both the backflow and the Jastrow terms are derived within the Random Phase Approximation (RPA) and are free of variational parameters\cite{Holzmann2003}. We further employ additional empirical functions for both the Jastrow and the backflow terms \cite{Kwon1998} which introduce a limited number of variational parameters (13). More details are provided in refs \cite{Pierleoni2006,Holzmann2003,Pierleoni2008,Morales2014entropy}. Optimization is only performed on a representative set of nuclear configurations generated during a preliminary CEIMC run with unoptimized parameters. This strategy has been found to introduce a negligible energy bias of $\sim0.02mH/atom$ in our simulations (see the SM section of ref.\cite{Pierleoni2016}).

The nuclear configuration space is sampled with a multi-level Metropolis scheme and the Penalty method \cite{Pierleoni2006}. Quantum nuclei within the Path Integral formalism are sampled with specialized procedures which made the method computationally affordable as explained in refs \cite{Pierleoni2016,Rillo2017}. Within  this method a few proton slices are enough to accurately simulate systems above T=600K in the relevant range of densities. For example, the transition pressure converged with 8 proton slices at T=600K, the lowest temperature we considered (see the SM of ref. \cite{Pierleoni2016}). Nuclear configuration space is visited by a Smart Monte Carlo method with global attempted moves using nuclear forces from DFT. 

\section{Results}
\begin{figure}
  \centering
  \includegraphics[width=0.9\columnwidth]{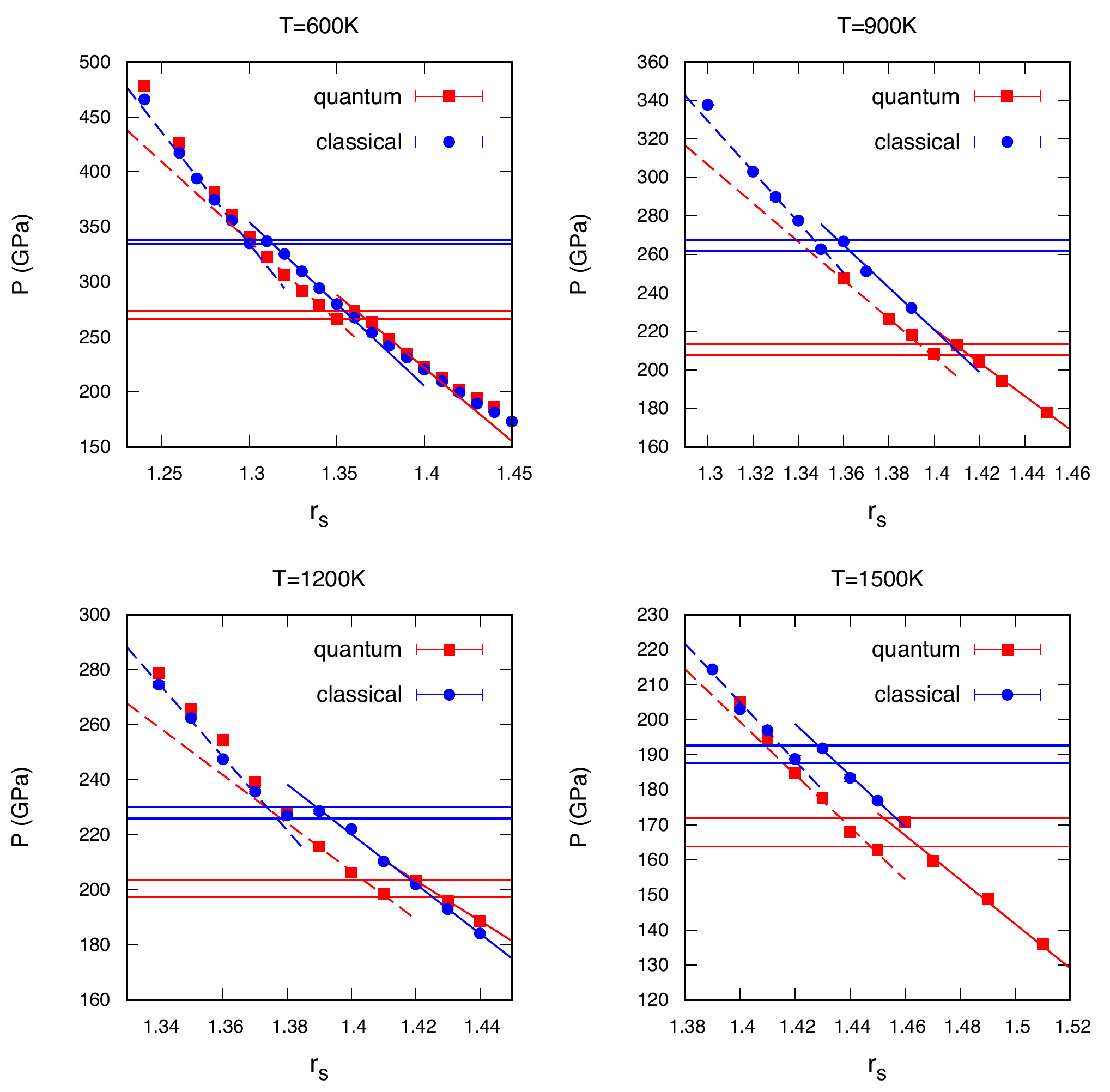}
  \caption{ Equation of state along four isotherms below the critical temperature for both quantum and classical nuclei. The specific volume is reported in terms of the coupling parameter $r_s$ which is related to the mass density in hydrogen by the relation $\rho(g/cm^3)=2.696/r_s^3$. The region of specific volume around the first order phase transition is illustrated for all data. For each system two horizontal lines indicate the coexistence region and the uncertainty on the detected transition pressure (related to the chosen density grid). For each system the two branches of the EOS near the transition points are fit by straight lines used to infer the compressibility.}
  \label{fig:EOS}
\end{figure}    
We perform simulations of systems of 54 protons and 54 electrons along five isotherms $T=600K, 900K, 1200K, 1500K, 3000K$. In figure \ref{fig:EOS} we report the equation of state  (EOS) along the four isotherms at lower temperature, where a discontinuity in the EOS is observed. Since our simulations are at constant volume, a first order phase transition manifests as the occurrence of two different branches of the EOS, one before and one after the transition pressure. The very weak first order character of this transition allows one to pinpoint rather precisely the transition pressure without recurring to the more demanding determination of the absolute free energies of the two different phases. We do not observe long-lived metastable states. In fig. \ref{fig:EOS} the horizontal lines indicate the transition pressure with its uncertainty obtained as the difference between the highest pressure observed for the low density phase and the lowest pressure observed for the high density phase. This determination of the transition pressure is somewhat sensitive to our density grid and might be improved by choosing a finer grid.
The data shown in fig \ref{fig:EOS} are from VMC and include finite size corrections as explained in the SM of ref. \cite{Pierleoni2016} (see also \cite{Holzmann2016}). The corresponding numerical values are given in the SM of ref. \cite{Pierleoni2016}.
\begin{figure}
  \centering
  \includegraphics[width=0.8\columnwidth]{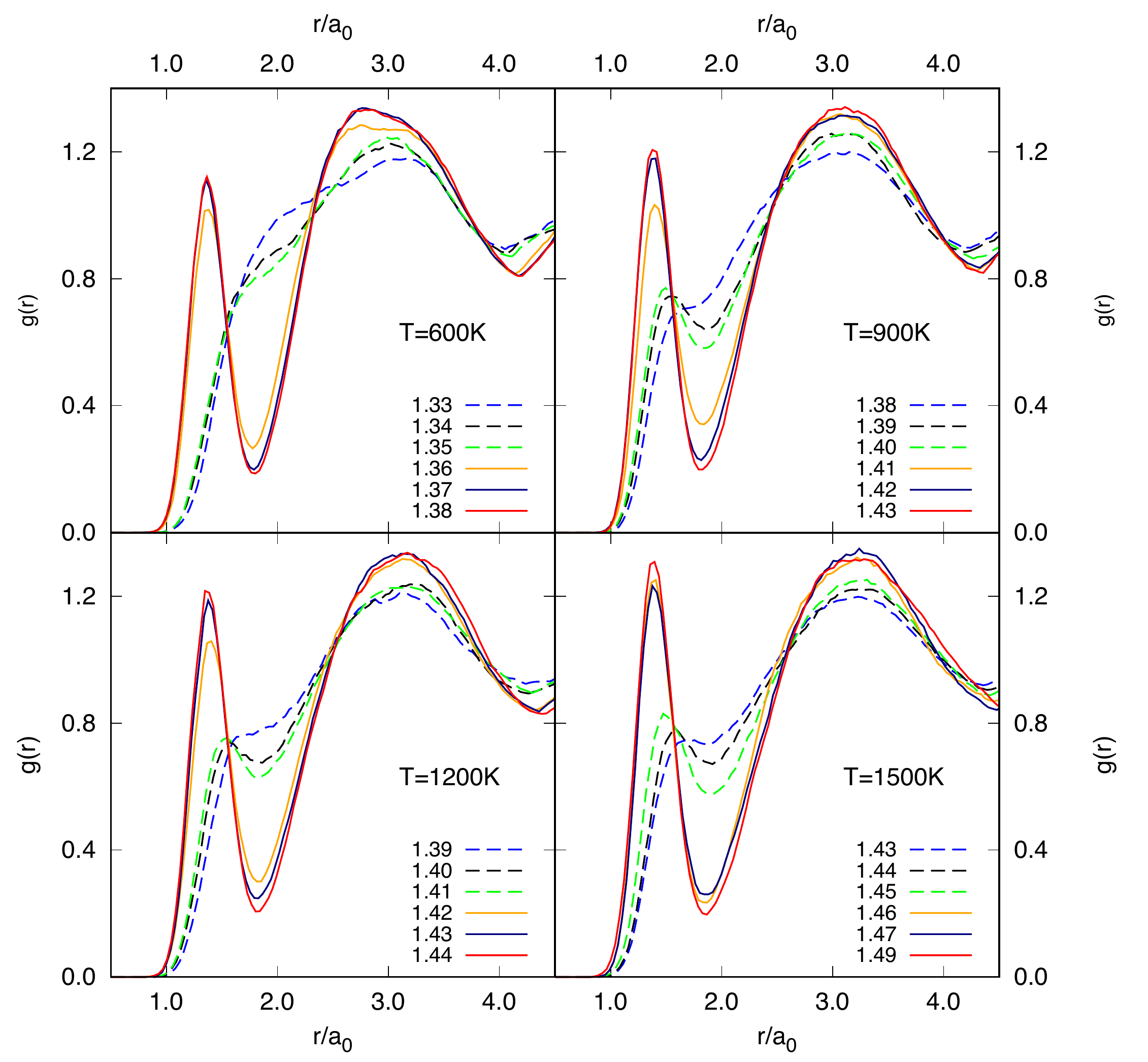}
  \caption{Proton-proton radial distribution function along four isotherms below the critical temperature for quantum nuclei. For sake of clarity we only show the three densities in each phase closest to the transition point, as indicated in the legend of each panel. The abrupt character of the molecular dissociation at the transition is clear. At each temperature the green and the orange lines correspond to the two coexisting densities.}
  \label{fig:gr_q}
\end{figure} 
In fig \ref{fig:gr_q} we report, along the same isotherms and for quantum protons only, the proton-proton radial distribution function for six different densities around the observed coexisting densities. We observe the sudden strong reduction or disappearance of the molecular peak precisely at the density discontinuity in the EOS. In each panel the green and orange lines represent the data for the two coexisting densities. Results for classical protons are similar. From figures \ref{fig:EOS} and \ref{fig:gr_q} it is evident that the thermodynamic transition corresponds to the molecular dissociation. In ref \cite{Pierleoni2016} we have  provided further evidence that a change in the electronic properties of the system occurs at the same transition point, turning the insulator molecular fluid into a conducting fluid. From fig. \ref{fig:gr_q} it appears that the conducting fluid is nearly monoatomic; we will provide more support for this in the next section. In this paper we do not discuss  the electronic properties of the fluid.
\begin{figure}
  \centering
  \includegraphics[width=\columnwidth]{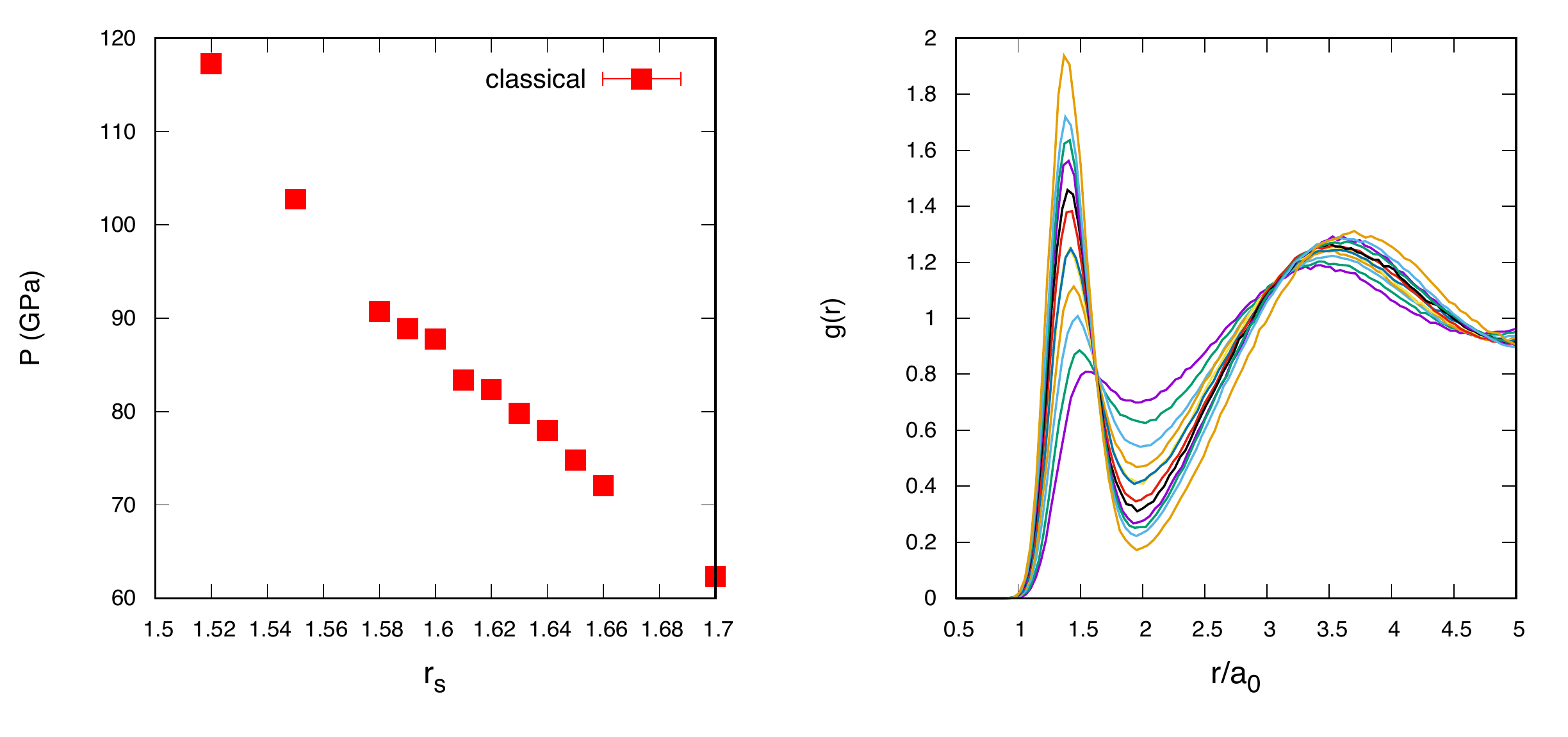}
  \caption{EOS (left panel) and proton-proton radial distribution function (right panel) along the T=3000K isotherm for classical protons only. Note the progressive character of the disappearance of the molecular peak.}
  \label{fig:3000K}
\end{figure} 
In fig. \ref{fig:3000K} we report both the EOS and the radial distribution function along the $T=3000K$ isotherm. At such high temperatures we have limited our investigation to classical protons since we expect nuclear quantum effects to be negligibly small. The EOS does not present a clear sign of discontinuity signaling that this temperature is above the critical point of the first order transition line. This is confirmed by the behavior of the $g(r)$ which exhibit a progressive decrease of the molecular peak without the discontinuous behavior observed at lower temperatures. A change in the slope of the EOS is apparent around $r_s\simeq 1.6$ which could be the signature of the Widom line, the continuation of the transition line in the region of homogenous fluid \cite{Simeoni2010}. However, our present density grid is too sparse and the statistical fluctuation of the data too large to allow for a convincing determination of the isothermal compressibility which should present a maximum along this line. 

\section{Cluster analysis}
\begin{figure}
  \centering
  \includegraphics[width=0.7\columnwidth]{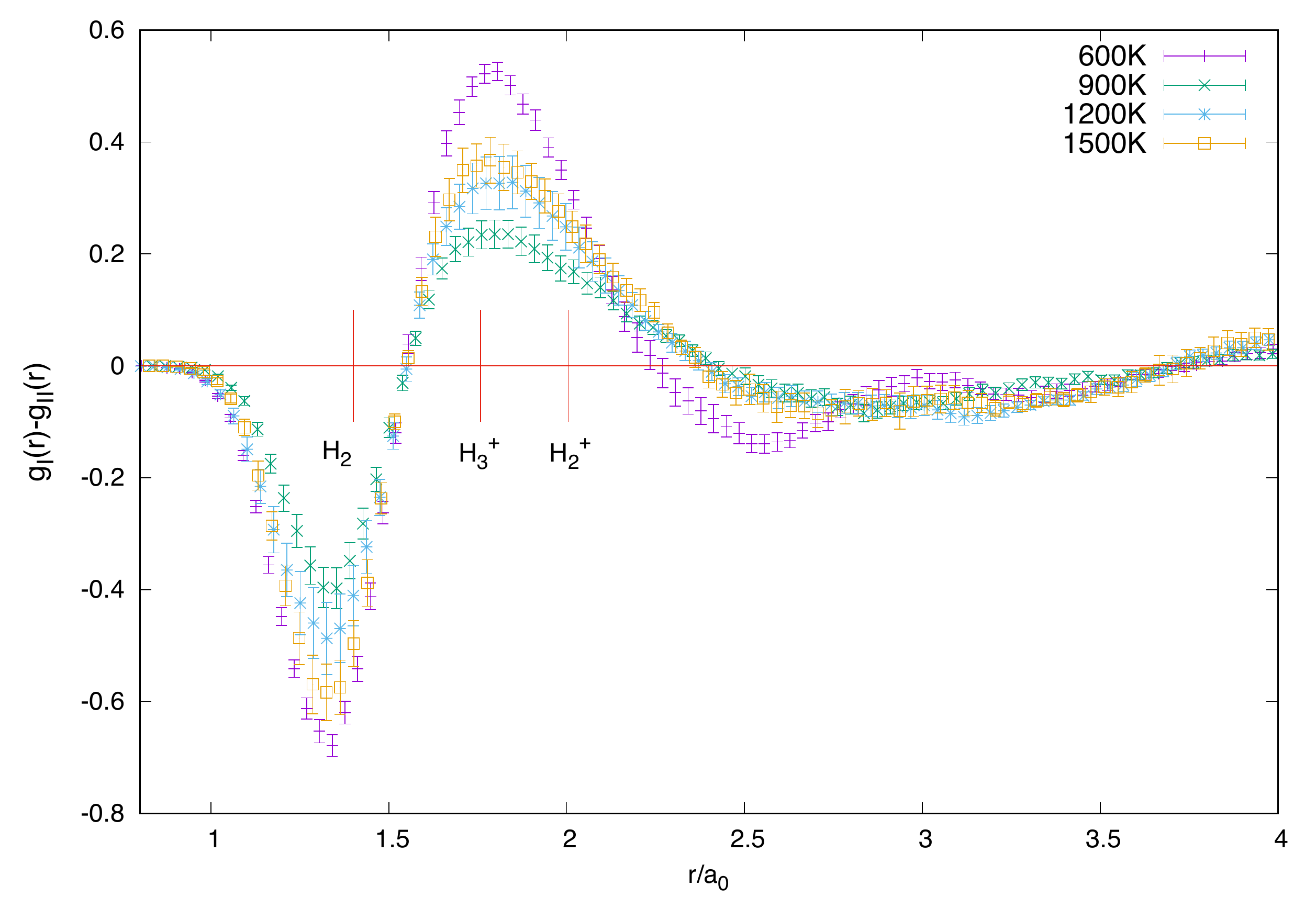}
  \caption{Difference between the radial distribution functions of the two coexisting densities $\Delta g(r)=g_{II}(r)-g_{I}(r)$ along the four isotherms below the critical point. Phase I is the molecular phase, while phase II is the dissociated phase beyond the transition. Typical distances in the isolated molecule and in the ions $H_2^+$ and $H_3^+$ are indicated by the vertical segments.}
  \label{fig:h3}
\end{figure}
Recently Norman and Saitov \cite{Norman2017a,Norman2017b} used BOMD with classical protons and the PBE functional, to study the same phase transition. Based on the comparative analysis of the $g(r)$'s of the two coexisting densities, they proposed that in the fluid just beyond the transition line, the molecular phase transforms to a mixture of $H^+$, $H_3^+$ and $H_2^+$. In figure \ref{fig:h3} we report the difference between the $g(r)$ of the two coexisting densities along the four isotherms below the critical point from our CEIMC calculations. The appearance of the maximum between $1.5 a_0$ and $2.4 a_0$ has been taken as evidence of the presence of those aggregates since their typical size in the gas phase is in this range of distances (see the vertical lines in the figure). 
In order to  characterize the transition microscopically and address the question of the nature of the fluid on the two sides of the transition line we performed a cluster analysis of the nuclear configurations sampled during the CEIMC runs. For each nuclear configuration we compute the table of all pair distances with a fixed cutoff, we sort it in ascending order based on the corresponding distance value, and we associate with each proton the closest proton not already involved in a previously found pair (i.e. corresponding to a smaller distance) and the two closest protons not already involved in previously found trimers. We performed two different analysis for two different purposes. The first analysis uses a cutoff distance of $1.8a_0$, corresponding to the position of the first minimum of the $g(r)$, and is exploited to estimate the distribution of ``molecules'' (namely the number of assigned pairs) and their average number in a way similar to the one in ref \cite{Vorberger2007a}. The second analysis uses a cutoff distance of $2.6a_0$, roughly the distance at which $\Delta g(r)=g_{II}(r)-g_{I}(r)$ vanishes, and is exploited to obtain a different measure of the molecular fraction and to investigate the stability of the trimers. Such large cutoff assigns each proton to a pair and to a trimer even for fully dissociated states, except in the case in which all protons in the first neighbor shell are already involved in previously found aggregates (corresponding to shorter bonds).
Finally a third estimate of the molecular fraction, namely twice the value of the coordination number\footnote{the coordination number $K(r)$ is defined as the three dimensional running integral of the radial distribution function multiplied by the particle density.} at $r=1.4a_0$, proposed in ref. \cite{Holst2008},  is exploited. 
\begin{figure*}
  \centering
  \includegraphics[width=0.9\textwidth]{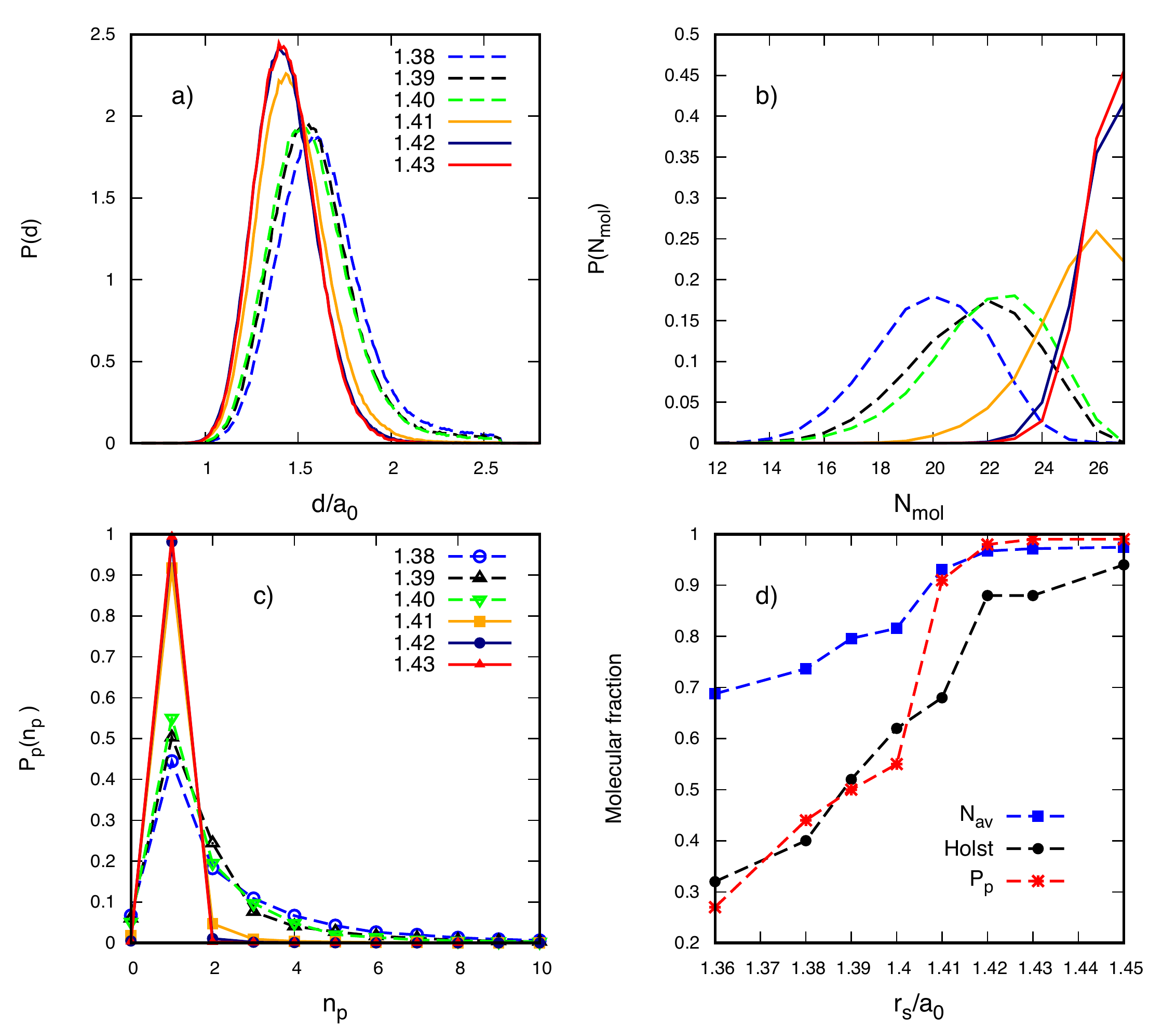}
  \caption{T=900K, quantum protons. Panel a): distribution of the distance in the pairs found by the cluster algorithm with the large cutoff of $2.6a_0$. Panel b) distribution of the number of found pairs with the short cutoff $1.8a_0$. Panel c) distribution of the number of different partners for a given protons in the pairs found with the large cutoff. Panel d) molecular fraction from different estimators: $N_{av}$ from the average number of molecules with the short cutoff, Holst from the coordination number at $r=1.4a_0$ and $P_p$ from the values of the distribution in panel c) for $n_p=1$.}
  \label{fig:t900_pairs}
\end{figure*} 

In principle these distance criteria have not much in common with the quantum mechanical picture of the chemical bond as the accumulation of the electronic pairs in singlet states around the center of the bond, but are instead based on the geometry of the configurations. However, in the BO picture there is a direct link between the geometry of the nuclear configurations and the distribution of electronic charge density.  Moreover, it is not clear how to characterize the nature of the dissociation process in a dense environment \cite{Riffet2017}. Nonetheless we find that clear indications of molecular dissociation are obtained by this simple cluster analysis with discontinuous changes of various observables. 
Figure \ref{fig:t900_pairs} shows the results of the analysis of pairs for the system of quantum nuclei along the T=900K isotherm for the same six densities around the transition pressure of figure \ref{fig:gr_q}. Results for the others cases (not shown) present similar features. Panel a) and c) reports results of the  analysis performed with the large cutoff. In panel a) the distribution of the distances averaged over all pairs and along the simulation runs is reported. We see a clear discontinuous change in this distribution at the transition. In the molecular phase I, the distribution is peaked at $\sim 1.4a_0$, it is rather narrow and does not depend on the density (the latter property is more clearly seen by comparing more densities, not shown for sake of clarity). Conversely in the dissociated phase II, the maximum of the distribution is at $\sim 1.6a_0$, the distribution gets more spread with a larger tail at large distances and depends on density. Note that the position of the maximum is far shorter than the equilibrium distance of $H_2^+$ ions. In panel c) we report the probability $P_p(n_p)$  for a single proton to belong to $n_p$ different pairs during the simulation run. Here also we see a striking change at the transition: in phase I, $P_p(n_p)$ is strongly peaked at $n_p=1$ with a very small value at $n_p=0$ and a vanishing small value for $n_p>1$. As before it is independent of the density. 
In phase II instead the distribution is still peaked at $n_p=1$ but with a lower value ($P_p(n_p=1)\simeq 0.5$), a decreasing exponential tail for $n_p>1$ and a value at the origin of $\sim 0.1$ meaning that 10\% of the protons are not paired by the cluster algorithm, despite the large cutoff. Moreover it presents a detectable dependence on density, becoming less structured for increasing density. In phase II at coexistence a proton has $\sim$10\% of probability to be unpaired, $\sim$50\% of probability to stay paired with the same partner, $\sim$20\% of probability to stay paired with a second partner, $\sim$10\% of probability to visit a third partner and so forth. Therefore we observe a large mobility of the bonding network completely absent in phase I even at coexistence.
\begin{figure*}
  \centering
  \includegraphics[width=0.9\textwidth]{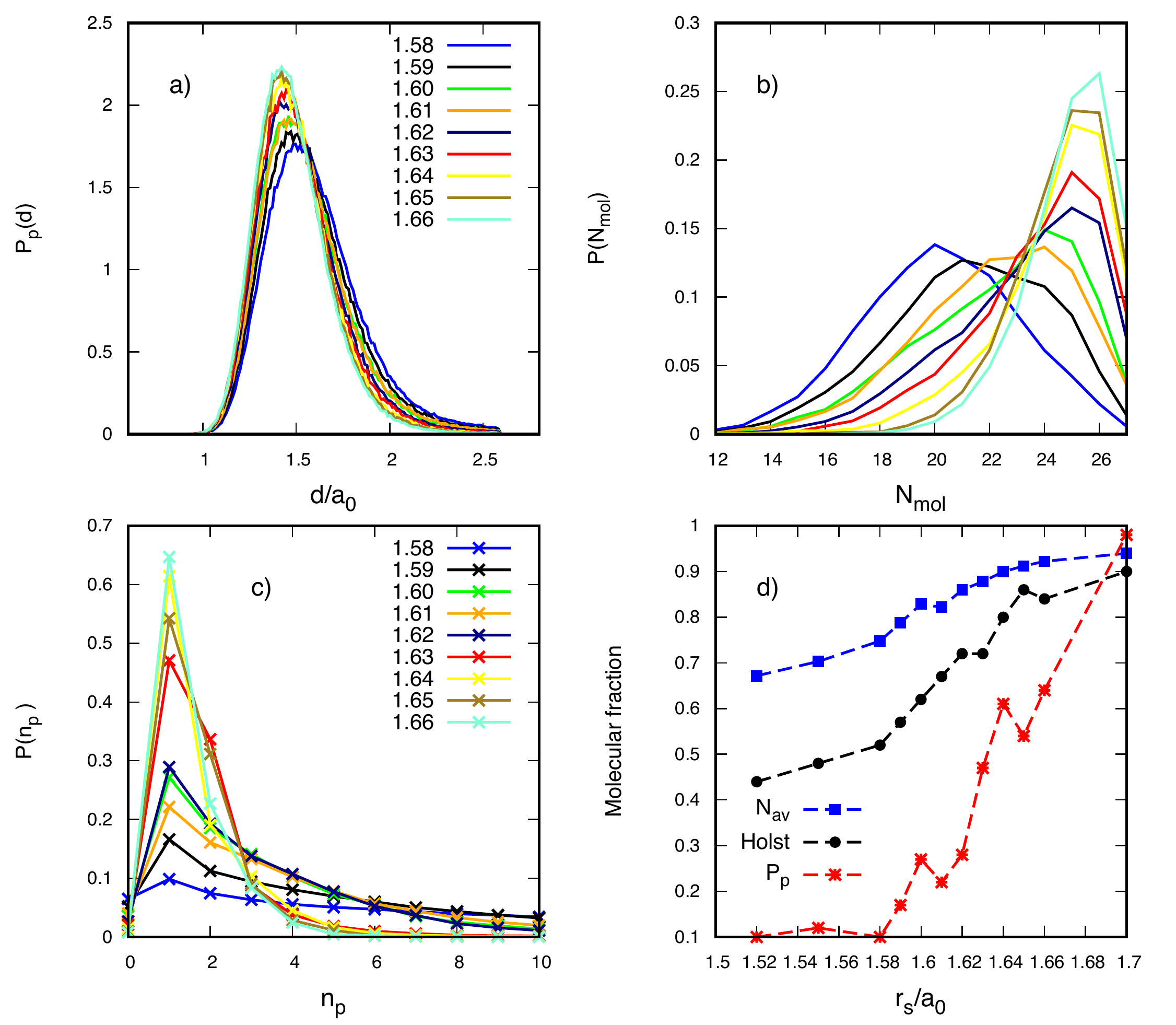}
  \caption{The panels report the same results as in figure \ref{fig:t900_pairs} but along the high temperature isotherm $T=3000K$.}
  \label{fig:t3000_pairs}
\end{figure*}    
Panel b) of figure \ref{fig:t900_pairs} reports the distribution of the number of detected ``molecules" using the short cutoff distance ($1.8a_0$). Again we see a discontinuous change in the form of the distribution at the transition. The average number of molecules from those distributions provides the first estimator of the molecular fraction. It is roughly equivalent to compute the coordination number at $r=1.8a_0$ of the difference between the total $g(r)$ and the $g(r)$ between non-bounded pairs. Results from this estimator, reported in panel d) of the same figure and denoted by $N_{av}$, completely mask the occurrence of the transition and suggest the presence of a large number of molecules even in phase II. In ref \cite{Vorberger2007a} this estimator was corrected by reducing the number of molecules based on a dynamical criterion: only bonded pairs that lasted more than some number (10) of oscillation periods were retained. This correction is difficult to apply for quantum nuclei since the simulation does not provide direct access to dynamical properties. Moreover we perform a Monte Carlo sampling (although with global moves) again not providing dynamical information. 
\begin{figure*}
  \centering
  \includegraphics[width=0.9\textwidth]{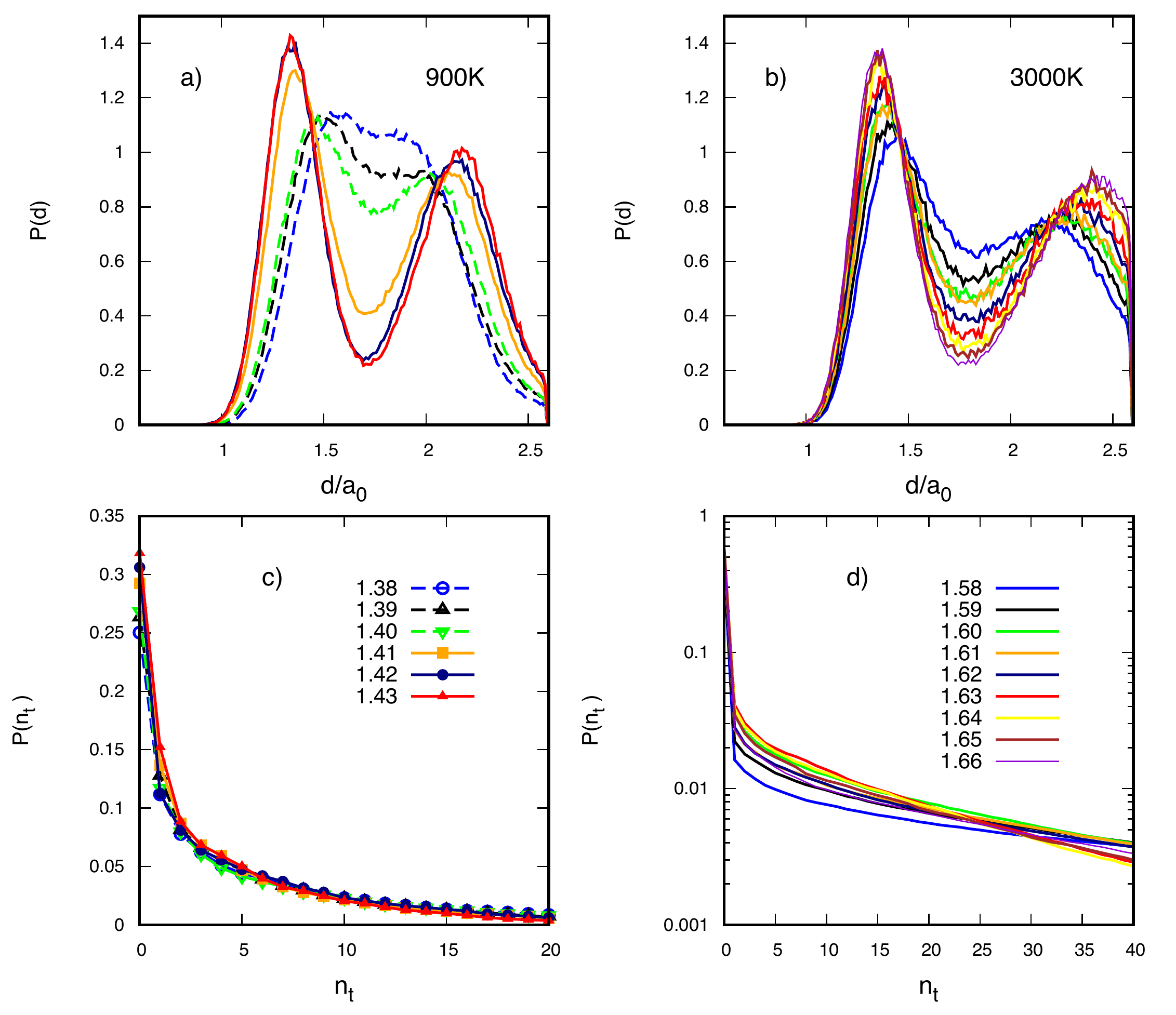}
  \caption{Analysis of triplets. Results for the isotherm at $T=900K$ are in the left column, while results at $T=3000K$ are in the right column. Panels a) and b): distribution of the two shortest distances within trimers. Panel c) and d) Probability for a proton to belong to be assigned to $n_t$ trimers during the simulation runs.}
  \label{fig:triplets}
\end{figure*}   
In the same panel d) we report the estimator based on the coordination number of the total $g(r)$ at $r=1.4a_0$ (denoted by Holst). This estimator provides a lower molecular fraction in the dissociated phase but it does not show a discontinuity at the transition and provides a lower molecular fraction in phase I. Finally the third estimator based on the maximum of the probability in panel c) (denoted by $P_p$) is also reported in panel d). This estimator, although not based on the sequentiality of pairing along the generated trajectory, attempts to include the concept of persistent pairing. It provides an almost completely molecular system in phase I, presents a large discontinuity at the transition, and a rapidly decreasing molecular fraction with increasing density in phase II. Interestingly this estimator is in agreement with the $N_{av}$ estimator in phase I, and with the Holst estimator in phase II. At coexistence it assigns a molecular fraction of 0.90 in phase I and of 0.55 in phase II. However it is difficult to judge the significance of the latter number. In phase II the distribution of bond distances in panel a) indicates an average distance larger than the typical bond length, but it could still result from mixing molecules and unbounded pairs. However, we could not find any pair lasting for the entire trajectory: even the observed maximum of $P(n_p)$ at $n_p=1$ results from a persistent but not permanent bonding. Therefore we believe that the new estimator proposed in this work is good in phase I but it only provides an upper bond for the molecular fraction in phase II. 

It is interesting to perform the same analysis along the T=3000K isotherm where no phase transition has been detected. In panel a) of figure \ref{fig:t3000_pairs} we see that the probability of distances for bonded pairs changes progressively with density.
A small discontinuity is seen in the $P(n_p)$ between $r_s=1.63$ and $r_s=1.62$ (panel c)), both in the value at $n_p=1$ and in the amplitude of the tail at larger values of $n_p$, but it does not seem to be associated with a discontinuity in the EOS or in the radial distribution function (see figure \ref{fig:3000K}). 
Also the distribution of the bonded pairs with the short cutoff, shown in panel b), change progressively with density as well as the various estimators for the molecular fraction, shown in panel d).  The $P_p$ estimator provides the lowest molecular fraction but again this should be considered as an upper bound only.  

Finally we discuss the result of the analysis of trimers illustrated by figure \ref{fig:triplets}. Panel a) and b) report the distribution of the two shortest distance within a trimer at T=900K and T=3000K respectively. Below the critical temperature (left column) the distribution of distances has again a discontinuous behavior. Before the transition, in phase I we have a well-characterized two peak structure which signals the presence of well-formed molecules and a larger distance between different molecules. After the transition in phase II, this structure becomes less well-defined and the distribution transforms to a single large peak embracing both previous peaks. Note that the typical distances in this phase are compatible with the formation of stable trimers as proposed by Norman and Saitov. To judge about the stability of those trimers we computed the probability for a proton to belong to different trimers. This probability for T=900K and T=3000K is shown in panel c) and panel b) respectively. Below the critical temperature this distribution exhibits no detectable sign of the transition. At both temperatures it decreases monotonously with a long tail which show the unstable character of the formed trimers.

In conclusion we have presented further data analysis of CEIMC simulations along five isotherms in the range 600K<T<3000K and in the range of densities suitable to detect the molecular dissociation. A previous account of these results, published in ref \cite{Pierleoni2016}, pointed to the existence of a weakly first order thermodynamic transition for $T\leq 1500K$ accompanied by a sudden disappearance of molecules and appearance of electrical conductivity and electron delocalization. In this paper we have presented a more extensive analysis of the local nuclear structure around the transition line providing data for the proton-proton pair correlation functions and a cluster analysis to determine the formation of other aggregates in the dissociated phase recently proposed\cite{Norman2017a,Norman2017b}. We indeed confirm the abrupt character of the transition on structural properties other than the EOS and the $g(r)$ but do not find evidence of the formation of stable aggregates such as $H_3^+$. 
At the highest temperature studied, T=3000K, no discontinuity is observed in the EOS and in the $g(r)$ showing a continuous dissociation. However, the cluster analysis suggests the possibility of a small discontinuity in the local structure of the fluid. 

C.P. was supported by the Agence Nationale de la Recherche (ANR) France, under the program ``Accueil de Chercheurs de Haut Niveau 2015'' project: HyLightExtreme. D.M.C. was supported by DOE Grant NA DE-NA0001789 and by the Fondation NanoSciences (Grenoble).  Computer time was provided by PRACE Projects 2013091918 and 2016143296 and by an allocation of the Blue Waters sustained petascale computing project, supported by the National Science Foundation (Award OCI 07- 25070) and the State of Illinois.  

\bibliography{dottorato}

\begin{bibdiv}
\begin{biblist}

\bib{McMahon2012a}{article}{
      author={McMahon, Jeffrey~M.},
      author={Morales, Miguel~A.},
      author={Pierleoni, Carlo},
      author={Ceperley, David~M.},
       title={{The properties of hydrogen and helium under extreme
  conditions}},
        date={2012},
        ISSN={0034-6861},
     journal={Rev. Mod. Phys.},
      volume={84},
       pages={1607\ndash 1653},
         url={http://link.aps.org/doi/10.1103/RevModPhys.84.1607},
}

\bib{Rillo2017}{article}{
      author={Rillo, Giovanni},
      author={Morales, Miguel~A.},
      author={Ceperley, David~M.},
      author={Pierleoni, Carlo},
       title={{Coupled Electron-Ion Monte Carlo simulation of hydrogen
  molecular crystals}},
        date={2017},
     journal={J. Chem. Phys. (in print)},
       pages={1\ndash 16},
      eprint={http://arxiv.org/abs/1708.07344},
         url={http://arxiv.org/abs/1708.07344},
}

\bib{Knudson2015}{article}{
      author={Knudson, M.~D.},
      author={Desjarlais, M.~P.},
      author={Becker, A.},
      author={Lemke, R.~W.},
      author={Cochrane, K.~R.},
      author={Savage, M.~E.},
      author={Bliss, D.~E.},
      author={Mattsson, T.~R.},
      author={Redmer, R.},
       title={{Direct observation of an abrupt insulator-to-metal transition in
  dense liquid deuterium}},
        date={2015},
        ISSN={0036-8075},
     journal={Science},
      volume={348},
       pages={1455\ndash 1460},
         url={http://www.sciencemag.org/cgi/doi/10.1126/science.aaa7471},
}

\bib{Ohta2015}{article}{
      author={Ohta, Kenji},
      author={Ichimaru, Kota},
      author={Einaga, Mari},
      author={Kawaguchi, Sho},
      author={Shimizu, Katsuya},
      author={Matsuoka, Takahiro},
      author={Hirao, Naohisa},
      author={Ohishi, Yasuo},
       title={{Phase boundary of hot dense fluid hydrogen}},
        date={2015},
        ISSN={2045-2322},
     journal={Scientific Reports},
      volume={5},
       pages={16560},
         url={http://dx.doi.org/10.1038/srep16560},
}

\bib{Zaghoo2016}{article}{
      author={Zaghoo, Mohamed},
      author={Salamat, Ashkan},
      author={Silvera, Isaac~F.},
       title={{Evidence of a first-order phase transition to metallic
  hydrogen}},
        date={2016},
        ISSN={1550235X},
     journal={Phys. Rev. B},
      volume={93},
      number={15},
       pages={155128},
}

\bib{Lorenzen2010}{article}{
      author={Lorenzen, W},
      author={Holst, B},
      author={Redmer, R},
       title={{First-order liquid-liquid phase transition in dense hydrogen}},
        date={2010},
        ISSN={1095-3795},
     journal={Phys. Rev. B},
      volume={82},
      number={19},
       pages={195107},
         url={papers://68dd8153-82a1-4a0e-9ae0-80ae65a14bf7/Paper/p1561},
}

\bib{Morales2010}{article}{
      author={Morales, Miguel~A},
      author={Pierleoni, Carlo},
      author={Schwegler, Eric},
      author={Ceperley, D~M},
       title={{Evidence for a first-order liquid-liquid transition in
  high-pressure hydrogen from ab initio simulations}},
        date={2010},
     journal={Proc. Nat. Acad. Sc.},
      volume={107},
      number={29},
       pages={12799\ndash 12803},
}

\bib{Morales2013liquid}{article}{
      author={Morales, Miguel~A.},
      author={McMahon, Jeffrey~M.},
      author={Pierleoni, Carlo},
      author={Ceperley, David~M.},
       title={{Nuclear quantum effects and nonlocal exchange-correlation
  functionals applied to liquid hydrogen at high pressure}},
        date={2013},
        ISSN={00319007},
     journal={Phys. Rev. Letts.},
      volume={110},
      number={6},
       pages={65702},
}

\bib{Pierleoni2016}{article}{
      author={Pierleoni, Carlo},
      author={Morales, Miguel~A},
      author={Rillo, Giovanni},
      author={Holzmann, Markus},
      author={Ceperley, David~M},
       title={{Liquid–liquid phase transition in hydrogen by coupled
  electron–ion Monte Carlo simulations}},
        date={2016},
        ISSN={0027-8424},
     journal={Proc. Natl. Acad. Sci.},
      volume={113},
      number={18},
       pages={4954\ndash 4957},
         url={http://www.pnas.org/lookup/doi/10.1073/pnas.1603853113},
}

\bib{Holst2008}{article}{
      author={Holst, Bastian},
      author={Redmer, Ronald},
      author={Desjarlais, Michael~P.},
       title={{Thermophysical properties of warm dense hydrogen using quantum
  molecular dynamics simulations}},
        date={2008},
        ISSN={10980121},
     journal={Phys. Rev. B},
      volume={77},
       pages={184201},
}

\bib{Vorberger2007a}{article}{
      author={Vorberger, J},
      author={Tamblyn, I},
      author={Militzer, B},
      author={Bonev, S},
       title={{Hydrogen-helium mixtures in the interiors of giant planets}},
        date={2007},
        ISSN={1095-3795},
     journal={Phys. Rev. B},
      volume={75},
      number={2},
       pages={24206},
         url={papers://68dd8153-82a1-4a0e-9ae0-80ae65a14bf7/Paper/p657},
}

\bib{Pierleoni2006}{incollection}{
      author={Pierleoni, C.},
      author={Ceperley, D.M.},
       title={{The coupled electron-ion monte carlo method}},
        date={2006},
   booktitle={Lecture notes in physics},
      volume={703},
   publisher={Springer, Berlin},
       pages={641\ndash 683},
}

\bib{Holzmann2003}{article}{
      author={Holzmann, M},
      author={Ceperley, D~M},
      author={Pierleoni, C},
      author={Esler, K},
       title={{Backflow correlations for the electron gas and metallic
  hydrogen.}},
        date={2003},
        ISSN={1063-651X},
     journal={Phys. Rev. E},
      volume={68},
       pages={046707},
}

\bib{Kwon1998}{article}{
      author={Kwon, Y},
      author={Ceperley, D},
      author={Martin, R},
       title={{Effects of backflow correlation in the three-dimensional
  electron gas: Quantum Monte Carlo study}},
        date={1998},
        ISSN={1095-3795},
     journal={Phys. Rev. B},
      volume={58},
      number={11},
       pages={6800\ndash 6806},
         url={papers://68dd8153-82a1-4a0e-9ae0-80ae65a14bf7/Paper/p1389},
}

\bib{Morales2014entropy}{article}{
      author={Morales, Miguel~A},
      author={Clay, Raymond},
      author={Pierleoni, Carlo},
      author={Ceperley, David~M},
       title={{First Principles Methods: A Perspective from Quantum Monte
  Carlo}},
        date={2014},
     journal={Entropy},
      volume={16},
       pages={287\ndash 321},
}

\bib{Pierleoni2008}{article}{
      author={Pierleoni, Carlo},
      author={Delaney, Kris~T.},
      author={Morales, Miguel~A.},
      author={Ceperley, David~M.},
      author={Holzmann, Markus},
       title={{Trial wave functions for high-pressure metallic hydrogen}},
        date={2008},
        ISSN={00104655},
     journal={Comp. Phys. Comms.},
      volume={179},
      number={1-3},
       pages={89\ndash 97},
}

\bib{Holzmann2016}{article}{
      author={Holzmann, M.},
      author={Clay, R.C.},
      author={Morales, M.~A.},
      author={Tubmann, N.~M.},
      author={Ceperley, D.~M.},
      author={Pierleoni, C.},
       title={{Theory of Finite Size Effects for Electronic Quantum Monte Carlo
  Calculations of Liquids and Solids}},
        date={2016},
        ISSN={2469-9950},
     journal={Phys. Rev. B},
      volume={94},
       pages={035126},
         url={http://link.aps.org/doi/10.1103/PhysRevB.94.035126},
}

\bib{Norman2017a}{article}{
      author={Norman, G~E},
      author={Saitov, I~M},
       title={{Ionization of Molecules at the Fluid-€"Fluid Phase Transition in
  Warm Dense Hydrogen}},
        date={2017},
     journal={Dokl. Phys.},
      volume={62},
      number={6},
       pages={284\ndash 288},
  url={https://link.springer.com/content/pdf/10.1134{\%}2FS1028335817060118.pdf},
}

\bib{Norman2017b}{article}{
      author={Norman, G~E},
      author={Saitov, I~M},
       title={{Critical Point and Mechanism of the Fluid-Fluid Phase Transition
  in Warm Dense Hydrogen}},
        date={2017},
     journal={Dokl. Phys.},
      volume={62},
      number={6},
       pages={294\ndash 298},
  url={https://link.springer.com/content/pdf/10.1134{\%}2FS1028335817060088.pdf},
}

\bib{Riffet2017}{article}{
      author={Riffet, Vanessa},
      author={Labet, Vanessa},
      author={Contreras-Garc{\'{i}}a, Julia},
       title={{A topological study of chemical bonds under pressure: solid
  hydrogen as a model case}},
        date={2017},
     journal={Phys. Chem. Chem. Phys.},
      volume={19},
       pages={26381\ndash 26395},
         url={http://pubs.rsc.org/en/content/articlepdf/2017/cp/c7cp04349j},
}

\bib{Simeoni2010}{article}{
      author={Simeoni, G. G.},
      author={Bryk, T.},
      author={Gorelli, F. A.},
      author={Krisch, M.},
      author={Ruocco, G.},
      author={Santoro, M.},
      author={Scopigno, T.}, 
       title={{The Widom line as the crossover between liquid-like and gas-like behaviour in
  supercritical fluids}},
        date={2010},
     journal={Nature Physics},
      volume={6},
       pages={503\ndash 507},
         url={http://www.nature.com/nphys/journal/v6/n7/full/nphys1683.html?foxtrotcallback=true},
}

\end{biblist}
\end{bibdiv}

\end{document}